
\def\X{{\cal{X}}}

\def\>{\rangle}
\def\<{\langle}
\def\vev#1{{\left\langle #1 \right\rangle}}

\def\VEv#1{{\biggl\langle #1 \biggr\rangle}}

\def\2{{\frac{1}{2}}}

\def\1{{{-1}}}

\def\bz{{\bar z}}
\def\bdry{{bdry}}
\def\Z{{{\cal Z}}}

\def\P{{\partial}}
\def\bP{{\bar\partial}}






\def\a{{\alpha}}
\def\b{{\beta}}

\def\d{{\delta}}

\def\t{{\theta}}

\def\r{{\rho}}
\def\s{{\sigma}}

\def\k{{\kappa}}

\def\S{{\cal S}}

\input harvmac
\def\frac#1#2{{#1\over#2}}

\font\ensX=msbm10
\font\ensVII=msbm7
\font\ensV=msbm5
\newfam\math
\textfont\math=\ensX \scriptfont\math=\ensVII \scriptscriptfont\math=\ensV
\def\ensemble{\fam\math\ensX}


\lref\AlekseevMC{ A.~Y.~Alekseev and V.~Schomerus, ``D-branes in the
WZW model,'' Phys.\ Rev.\ D {\bf 60}, 061901 (1999)
[arXiv:hep-th/9812193].
}
\lref\GawedzkiBQ{ K.~Gawedzki, ``Conformal field theory: A case
study,'' arXiv:hep-th/9904145.
}
\lref\GawedzkiYE{ K.~Gawedzki, ``Boundary WZW, G/H, G/G and CS
theories,'' Annales Henri Poincare {\bf 3}, 847 (2002)
[arXiv:hep-th/0108044].
}
\lref\GerasimovGA{ A.~A.~Gerasimov and S.~L.~Shatashvili, ``Stringy
Higgs mechanism and the fate of open strings,'' JHEP {\bf 0101}, 019
(2001) [arXiv:hep-th/0011009].
}

\lref\GerasimovZP{
A.~A.~Gerasimov and S.~L.~Shatashvili,
``On exact tachyon potential in open string field theory,''
JHEP {\bf 0010}, 034 (2000)
[arXiv:hep-th/0009103].
}

\lref\KlimcikHP{ C.~Klimcik and P.~Severa, ``Open strings and
D-branes in WZNW models,'' Nucl.\ Phys.\ B {\bf 488}, 653 (1997)
[arXiv:hep-th/9609112].
}
\lref\PolyakovTT{ A.~M.~Polyakov and P.~B.~Wiegmann, ``Theory Of
Nonabelian Goldstone Bosons In Two Dimensions,'' Phys.\ Lett.\ B
{\bf 131}, 121 (1983).
}

\lref\SenRG{ A.~Sen, ``Stable non-BPS states in string theory,''
JHEP {\bf 9806}, 007 (1998) [arXiv:hep-th/9803194].
}
\lref\SenMG{ A.~Sen, ``Non-BPS states and branes in string theory,''
arXiv:hep-th/9904207.
}
\lref\SenIN{ A.~Sen, ``Tachyon matter,'' JHEP {\bf 0207}, 065 (2002)
[arXiv:hep-th/0203265].
}
\lref\ShatashviliKK{ S.~L.~Shatashvili, ``Comment on the background
independent open string theory,'' Phys.\ Lett.\ B {\bf 311}, 83
(1993) [arXiv:hep-th/9303143].
}
\lref\ShatashviliPS{ S.~L.~Shatashvili, ``On the problems with
background independence in string theory,'' Alg.\ Anal.\  {\bf 6},
215 (1994) [arXiv:hep-th/9311177].
}
\lref\StanciuID{ S.~Stanciu, ``D-branes in group manifolds,'' JHEP
{\bf 0001}, 025 (2000) [arXiv:hep-th/9909163].
}
\lref\StromingerZD{ A.~Strominger, ``Closed Strings In Open String
Field Theory,'' Phys.\ Rev.\ Lett.\  {\bf 58}, 629 (1987).
}
\lref\WittenQY{ E.~Witten, ``On background independent open string
field theory,'' Phys.\ Rev.\ D {\bf 46}, 5467 (1992)
[arXiv:hep-th/9208027].
}
\lref\WittenCR{ E.~Witten, ``Some computations in background
independent off-shell string theory,'' Phys.\ Rev.\ D {\bf 47}, 3405
(1993) [arXiv:hep-th/9210065].
} \lref\ZwiebachFE{ B.~Zwiebach, ``Oriented open-closed string
theory revisited,'' Annals Phys.\  {\bf 267}, 193 (1998)
[arXiv:hep-th/9705241].
} \lref\sam{S. Shatashvili, `` Closed strings as solitons in
background independent open string field theory'', unpublished, talk
at IHES, Paris, July 1997.}

\lref\BachasSY{ C.~Bachas and M.~Gaberdiel,
``Loop operators and the Kondo problem,''
arXiv:hep-th/0411067.
}

\lref\AFS{
A.~Alekseev, L.~D.~Faddeev and S.~L.~Shatashvili,
``Quantization Of Symplectic Orbits Of Compact Lie Groups By Means Of The
Functional Integral,''
J.\ Geom.\ Phys.\  {\bf 5}, 391 (1988).
}

\lref\orbitI{
A.~Alekseev and S.~L.~Shatashvili,
``Path Integral Quantization Of The Coadjoint Orbits Of The Virasoro Group And
2-D Gravity,''
Nucl.\ Phys.\ B {\bf 323}, 719 (1989).
}

\lref\orbitII{
A.~Alekseev and S.~L.~Shatashvili,
``Quantum Groups And WZW Models,''
Commun.\ Math.\ Phys.\  {\bf 133}, 353 (1990).
}

\lref\KMMI{
D.~Kutasov, M.~Marino and G.~W.~Moore,
``Remarks on tachyon condensation in superstring field theory,''
arXiv:hep-th/0010108.
}

\lref\KMMII{
D.~Kutasov, M.~Marino and G.~W.~Moore,
``Some exact results on tachyon condensation in string field theory,''
JHEP {\bf 0010}, 045 (2000)
[arXiv:hep-th/0009148].
}

\lref\fs{
L.~D.~Faddeev and S.~L.~Shatashvili,
``Algebraic And Hamiltonian Methods In The Theory Of Nonabelian Anomalies,''
Theor.\ Math.\ Phys.\  {\bf 60}, 770 (1985)
[Teor.\ Mat.\ Fiz.\  {\bf 60}, 206 (1984)].
}

\lref\michI{
J.~Mickelsson,
``On The 2 Cocycle Of A Kac-Moody Group,''
Phys.\ Rev.\ Lett.\  {\bf 55}, 2099 (1985).
}

\lref\michII{
J.~Mickelsson,
``Current Algebras And Groups,''
New York, USA: Plenum (1989) 313p.
}

\lref\lmns{
A.~Losev, G.~W.~Moore, N.~Nekrasov and S.~L.~Shatashvili,
``Central Extensions of Gauge Groups Revisited,''
arXiv:hep-th/9511185.
}

\lref\SeibergVS{
N.~Seiberg and E.~Witten,
``String theory and noncommutative geometry,''
JHEP {\bf 9909}, 032 (1999)
[arXiv:hep-th/9908142].
}

\lref\KontsevichVB{
M.~Kontsevich,
``Deformation quantization of Poisson manifolds, I,''
Lett.\ Math.\ Phys.\  {\bf 66}, 157 (2003)
[arXiv:q-alg/9709040].
}

\lref\PressleyQK{
A.~Pressley and G.~Segal,
``Loop Groups,''
Oxford, UK: Clarendon (1988) 318p. (Oxford Mathematical Monographies)
}

\lref\Shatashstrings{S. Shatashvili, On Field Theory of Open Strings, Tachyon Condensation and Closed Strings,
arXiv:hep-th/0105076, talk at String 2001, Mumbai, India, January 2001.}

\lref\SenNF{
A.~Sen,
``Tachyon dynamics in open string theory,''
arXiv:hep-th/0410103.
}

\lref\FigueroaOFarrillKZ{
J.~M.~Figueroa-O'Farrill and S.~Stanciu,
``D-brane charge, flux quantization and relative (co)homology,''
JHEP {\bf 0101}, 006 (2001)
[arXiv:hep-th/0008038].
}

\lref\AlekseevBS{
A.~Y.~Alekseev, A.~Recknagel and V.~Schomerus,
``Non-commutative world-volume geometries: Branes on SU(2) and fuzzy
spheres,''
JHEP {\bf 9909}, 023 (1999)
[arXiv:hep-th/9908040].
}

\Title{ \vbox{\baselineskip12pt
\hbox{HMI-04-02}
\hbox{TCD-MATH-04-23}
 }} {\vbox{
\centerline{Factorization Conjecture}
\vskip 4pt
\centerline{and the Open/Closed String Correspondence}
\bigskip
 }}
\medskip
\centerline{\bf Marcus Baumgartl$^{1}$, Ivo Sachs$^{1}$ and Samson L.
Shatashvili $^{2,3,4}$}
\vskip 0.5cm \centerline{\it $^{1}$ Theoretische Physik,
Ludwig-Maximilians Universit\"{a}t,}
\centerline{\it
Theresienstrasse 37,
D-80333, M\"{u}nchen, Germany}
\centerline{\it
$^{2}$ Department of Pure and Applied Mathematics, Trinity
College, Dublin 2, Ireland } \centerline{\it $^3$ Hamilton
Mathematics Institute, TCD, Dublin 2, Ireland}
\centerline{\it $^{4}$ IHES, 35 route de Chartres,
Bures-sur-Yvette, FRANCE}
\vskip 1cm
We present evidence for the factorization of the world-sheet path integrals for
2d conformal field theories on the disk into bulk and boundary
contributions.  This factorization is then used to
reinterpret a shift in closed string backgrounds in terms of boundary
deformations in background independent open string field
theory.
We give a proof of the factorization conjecture in the 
cases where the background is represented by WZW and
related models.

\medskip
\noindent \Date{December 2004}


\newsec{Introduction}

Since the early days of string theory it has been suspected that the
distinction between open strings and closed strings should not be
fundamental. This follows already from the observation that closed
string poles occur as intermediate states in open string scattering
amplitudes. From the point of view of open string field theory these poles
seem to violate unitarity unless closed string states are present in
the classical open string field theory. One possibility is to accept that
open string field theory is not unitary and to add extra closed
string degrees of freedom by hand \ZwiebachFE. However, in this
approach one has to address the problem of overcounting since now
the same diagram can be obtained from the open and closed string
sector of the field theory Lagrangian.

An alternative approach is to try to identify closed string states
directly in open string field theory \refs{\StromingerZD, \sam}. This idea
receives further motivation from Sen's work on non-BPS branes
\refs{\SenRG, \SenMG} which resulted in a very active study of open
string field theory in different formulations and some progress in
understanding the vacuum structure of open strings has been
achieved (see \SenNF\ for a review).

In this paper we suggest an approach based on the idea that closed
string degrees of freedom correspond to (non-local) boundary
interactions advocated in \refs{\GerasimovGA, \Shatashstrings}
within the framework of background independent open string
field theory (BSFT)
\refs{\GerasimovGA, \WittenQY\WittenCR\ShatashviliKK\ShatashviliPS\GerasimovZP\KMMI{--}\KMMII}.
BSFT is defined by the path integral over $\sigma$-model fields for a 
fixed closed string background $\X$ with the dynamical open string 
degrees of freedom $t$ corresponding to
boundary deformations of the CFT on the disk.  
It is, however, important to note that these deformations are not 
required to be local on the boundary of the world-sheet
\refs{\WittenQY\WittenCR\ShatashviliKK{--}\ShatashviliPS}.
Once non-local boundary perturbations are included the 
distinction to open and closed degrees of freedom on a 
world-sheet with boundary becomes ambiguous.
In fact in the early days of background independent open string
theory it was realized that the notion of locality on the world-sheet was a
major question to be addressed since deformations on the boundary were
described by a limiting procedure of taking the closed string
operator from the bulk and moving it to the boundary. The simplest
way to identify $\X$ is by means of the closed string $\sigma$-model
Lagrangian. $\X$ then defines a conformal $\sigma$-model background in
the absence of boundaries.
In the examples studied in this paper we will make some
natural choices in this regard and then demonstrate a relation between
them.

The key ingredient in our approach is the factorization property for the
BSFT space-time action $ \S(\X| t)$,
\eqn\EQNFactConj {
    \S(\X| t) = \Z_0(\X) \S_0(\X|t).}
Here
$\Z_0(\X)$ is the D-instanton partition function and $\S_0(\X|t)$ is
described purely in terms of the quantum mechanical degrees of
freedom $\Phi_b$ on the boundary. 
Given the relation between 
the world-sheet partition function and the BSFT
action (see equation (2.1) below)
this property is a consequence of the 
following 
conjecture for 2d conformal field theories on a manifold with
boundary:
\eqn\ant{\int_{\Phi|_{\partial \Sigma}=
\Phi_b(\theta)} D[\Phi]\, e^{{i \over {\alpha'}}I_{\Sigma}(\Phi)}=
\Z_0(\X)\,e^{{i \over {\alpha'}}I^\bdry(\Phi_b)}.}
Here $I_{\Sigma}(\Phi)$ is a
Lagrangian for the world-sheet conformal field theory on a 2d surface with
boundary, and $\Z_0(\X)$ is the D-instanton partition function 
which is given by \ant\ for $\Phi_b=0$.  We verify this
property the case of when $\Sigma$ is a disk, in the
situation where ghosts and matter decouple and for $\X$ such that
the  closed string world-sheet is conformal and described in terms
of a WZW (or related) model. These technical assumptions are
necessary since not much is known about BSFT when ghosts and matter
do not decouple.

The logic underlying our approach is the following: To each 2d CFT
with  boundary corresponds a boundary action $I^\bdry(\Phi_b)$. Due
to the factorization property, $I^\bdry(\Phi_b)$ is independent of
$\alpha'$ but certainly depends on the CFT chosen on the left hand
side of eqn. \ant.
The ambiguity in this process is under control (see section 3). 
On the other
hand, in the reconstruction of the bulk CFT for a given boundary action
there may be further ambiguities.  
We then claim that there is a distinction
between the class of bulk theories reconstructed from boundary
actions $I^\bdry$, differing by (non-local) functionals of the
boundary field $\Phi_b$. 

Note that 
the boundary action plays a central role
in BSFT since one integrates over
all maps from the world-sheet to the target space
without specifying the boundary conditions.
One starts from a boundary action and considers the
class of its boundary deformations; this class contains all other
boundary actions with the same number of boundary fields $\Phi_b$ (or
less).
The boundary actions corresponding to boundary conformal field theories on
the world-sheet are, by definition of the string field theory action, 
solutions of the classical equations of motion for $\S(\X|t)$. 
These are in turn critical points in $t$
for fixed $\X$ and denoted by $t_*$. The space-time action
$\S(\X|t_*+t_q)$ expanded around $t_*$ to $n$-th order in
$t_q$ is supposed to reproduce the $n$-point open string
amplitudes for the background defined by
$t_*$. This is known to be true on classical
level in the space-time field theory
corresponding to disk amplitudes on the world-sheet.

Concretely we start with a closed string background $\X$ and find $t_*^1$.
Then we look for a second critical
point of $\S_0$: $t^2_*$; since $\X$ is a ``hidden variable''
in the open string field theory action $\S_0$ we need to reconstruct
it for the new critical point $t_*^2$.
This in general is a difficult problem and in principle might be ambiguous. Even so
we can argue that in the set of critical points of
$\S_0(\X| t)$ there are critical points $t=t_*^1$ and $t=t_*^2$  such
that the expansion around $t_*^2$ is identical to the expansion around
$t_*^1$ but for different closed string background $\X'$, i.e.
\eqn\EQNConjecture {
    \S_0(\X|t_*^1+t_q) = \S_0(\X'|t_*^2+t_q).
}
Thus, a deformation from $t_*^1$ to $t_*^2$ can be interpreted as deforming the closed string
background from $\X$ to $\X'$.

A simple realization of the conjectured property \EQNConjecture\ leads to
the Seiberg-Witten map \SeibergVS: It is well known that a constant Kalb-Ramond
$B$-field can be seen equivalently as a closed string background $\X$ or a
perturbation on the boundary
of the open string world-sheet, i.e.
$\S_0(\X=(G,B)|t_q) = \S_0(\X'=(G,0)|t_*+t_q)$.
The result of \SeibergVS\ can then be formulated as the 
statement that the expansion around $t_*$ leads to 
non-commutative field theory in Minkowski space.
The generalization to a non-constant,
closed $B$-field leads to Kontsevich's deformation 
quantization \KontsevichVB. 
At present we allow for arbitrary $B$ compatible with bulk conformal invariance.

Note that the factorization of the world-sheet partition function
into bulk and boundary contributions is crucial for the closed
string degrees of freedom to be contained in open string 
field theory.
Indeed if bulk $\alpha'$-corrections entered in the definition of
$I^\bdry(\Phi_b)$ one would get different $\alpha'$-expansions for the open
and closed string $\beta$-function. 
The factorization property, which guaratees that closed string fluctuations 
do not feed back into the definition of the open string field theory,
is instrumental for the open-closed string correspondence to work. 
This appears to be a very subtle distinction
between bulk
conformal field theories in 2d and general 2d QFT where this factorization does not hold in
general.

As a warm up and to fix the notation, we apply our formalism in
section 2 to radius deformations on the torus where the decoupling
is immediate. As an example for a curved closed string background we
then prove the factorization property for boundary WZW models with
arbitrary boundary conditions to all orders in perturbation theory
in section 3. This requires a definition of WZW models with boundary
conditions which are not of the class $J=R\bar J$
\refs{\KlimcikHP\AlekseevMC\StanciuID\AlekseevBS{--}\FigueroaOFarrillKZ}, rather only implying
$T-\bar T=\beta^i(t)V_i(t)=0$, where $V_i(t)$ is a boundary
perturbation and $\beta^i(t)$ its $\beta$-function. Finally we
illustrate our result considering the SU(2)-WZW model in the large
radius limit in section 4.

\newsec{BSFT on a Torus}
\seclab\SecBSFTonTorus

In the case when ghost and matter fields decouple the definition of
the space-time action in flat space is written in terms of the disk
partition function $\Z(t)$ and boundary $\beta$-function as
\ShatashviliPS\foot{Note that this expression is written without use of a metric on the
space of boundary interactions, only the vector field $\beta^i$ is
required.}
\eqn\EBSFT{{\cal S}(\X|t) =
\left(1-\beta^i\frac{\P}{\P t^i}\right)\Z (\X|t).} Here $t^i$ are
the couplings representing the open string degrees of freedom and
$\beta^i$ denotes the $\beta$-function associated to the coupling
$t^i$.

For our purpose we suggest a different normalization of the
space-time action by replacing $\Z(\X|t)$ by $\Z^\bdry(\X|t)\equiv
\Z(\X|t)/\Z_0(\X)$, where $\Z_0(\X)$ is the ``D-instanton" partition
function, which is independent of the open string background
$\{t\}$. Of course this normalization assumes the factorization of
the CFT on the disk, which we will prove shortly.
Our normalization
does not alter the dynamics of the open string fields $t^i$,
therefore we can work with $\S_0(\X|t)$ instead of $\S(\X|t)$,

\eqn\EBSFTnew{{\cal S}_0(\X|t) =
\left(1-\beta^i\frac{\P}{\P t^i}\right)\Z^\bdry (\X|t).}

To start with we consider the free action for maps $X$ from the disk
into a circle of radius $R$
\eqn\Efreeaction{
    S(X) = \frac{R^2}{4\pi i\a'}\int_D \P X\bP X.
}
where $\P \equiv dz\P_z$. The radius $R$ plays the role of a
closed string modulus. According to BSFT we are instructed to
integrate over maps with free boundary conditions, which leads to the
notion of the boundary field $f$ defined through $X(z,\bar z)|_{\P
D} = f(\t)$; boundary deformations are functionals of $f$, in
general non-local. This field $f$ can be unambiguously extended from
the boundary to the interior of the disk via harmonicity condition
(harmonic functions are solutions of the world-sheet equations of
motion). Every field $X(z, \bar z)$ may thus be split into a
harmonic boundary field and a bulk field which obeys Dirichlet
conditions,

\eqn\expens{
    X(z, \bar z) = X_0(z, \bar z) + X_b(f),}
such that $X_0|_{\P D} = 0$ and $X_b(z, \bar z)|_{\P D} = f(\t)$
with $\Delta X_b = 0$, so $X_b(f)$ is a harmonic function with value
$f(\theta)$ on the boundary.

Note that the boundary field can always be expanded as $f=\sum_{n}
f_n e^{in\t}$, which suggests a separation into chiral and
anti-chiral modes corresponding to positive and negative
frequencies. Thus, $f = f^+ + f^- + f_0$ can then be extended to $X_b(f) =
f^+(z) + f^-(\bz) + f_0$. Moreover there is a reality condition ${f^+}^* =
f^-$. The zero mode $f_0$ plays the role of the space-time
integration variable in the space-time action.

Plugging this ansatz into the free action \Efreeaction\ the mixed
terms containing $X_0$ and $X_b$ vanish after partial integration.
The action splits into \eqn\Epolyakovbb{
    S(X) = \frac{R^2}{4\pi i\a'}\int \P X_0\bP X_0 + \frac{R^2}{4\pi i\a'}\int
\P f^+ \bP f^-. }
Given the translation invariance of the measure in this example
the factorization property is obviously satisfied. The partition function then reads
\eqn\EQNpartfnI{
    \Z(R) = \Z_0(R)\int D[f]\, e^{-\frac{R^2}{4\pi i\a'}\int \P f^+ \bP f^-},
}
where
\eqn\EQNpartfnII{
    \Z_0(R) = \int D[X_0]\, e^{-\frac{R^2}{4\pi i\a'}\int \P X_0\bP X_0},
    }
supplemented by the $b,c$ ghost system is the ``D-instanton''
partition function\foot{ Here we take the conventional boundary
conditions for $b$ and $c$, because decoupling of matter and ghost
sector is assumed. }. Since $f^\pm$ is harmonic its contribution
takes the form of a non-local boundary interaction
\eqn\hil{
    I^\bdry(f) = \frac{R^2}{4\pi}\oint f H(f)
        = \frac{R^2}{4\pi} \oint \oint d\t d\t' f(\t) H(\t,\t') f(\t')=
        \frac{R^2}{2\pi}\oint f^+ \partial_{\theta}f^-,
}
where $H$ is a Hilbert transform $H(f)=\P_n f = \partial (f_- -  f_+)$ and
$H(\t,\t')=\frac{1}{4\pi i}$$\sum_{n}$$e^{in(\t-\t')}|n|$. Integration over $f$ with
this boundary interaction then produces the partition function of a
D-brane extended along the $X$-direction.

To be more general we can add local interactions on the boundary, parametrized by couplings $\{t_q\}$.
They are given by
functionals of $f$, so that the local and non-local contributions can be collected into

\eqn\EQNbdryint{
    I^\bdry(t, X) = I^\bdry(t, f).
}
\eqn\boundarypar{
    \Z^\bdry(R|t)= \int D[f] e^{\frac{i}{\a'}I^\bdry(t,f)}.
}
>From the above it is now clear that a
change in the closed string modulus $R\to R+\d R$ appears as a deformation
of the boundary interaction
\eqn\addbndryint{\eqalign{
    I^\bdry &\to I^\bdry + \d I^\bdry\cr
    \d I^\bdry(R) &= \frac{R\d R}{2\pi}\oint f H(f).
}}
In the presence of open string degrees of freedom
this is a non-trivial modification of the boundary theory.
For instance  for the Euclidean D1-brane wrapping $S^1$
the condition for marginality of the boundary operator $\exp ikX_b$ with
$k=n\in {\ensemble Z}$ is changed by the shift\foot{Similarly,
strings attached to the D-instanton can wind around $S^1$. Their contribution to
the boundary partition function
is represented by the insertion of boundary vertex operators
$\exp i\frac{wR}{\a'}X_b$.} $R\to R+\delta R$.
We thus conclude that
the modulus $R$ of the closed string background
$\X=S^1_R$ enters as a non-local boundary interaction.
In particular,
\eqn\EQNShiftI{
    \eqalign{
    \S_0(\X=S^1_{R+\d R}| t^1_*+t_q)
    &=  \S_0(\X=S^1_R|t^2_* + t_q),
}}
in accord with \EQNConjecture.
Note that the theory without additional boundary interactions
is conformally invariant for any $R$.
Therefore there is no $\beta$-function associated to the radius deformation.
But the $\beta$-functions for other couplings depend on the non-local part
(and therefore on the bulk moduli) of the boundary interaction.

After this warm-up we will now consider interacting CFTs.
In the next section we show
that the factorization property also holds for boundary conformal
theories on group manifolds.

\newsec{Boundary WZW model}
\seclab\SecbWZW

The prototype example for open strings propagating in curved
space-time is the WZW model which is also an example where the $B$-field is not closed. Here we will discuss this case in
detail.
Other curved target spaces can be treated in a similar fashion.

As is well known  \refs{\KlimcikHP\AlekseevMC{--}\StanciuID} in this case
world-sheets $\Sigma$ with boundary $\P\Sigma$  require some care in
the definition of the topological term $\Gamma(g)=
\int_{\Gamma}\tr(dg g^{-1})^3$ with $\partial\Gamma=M$. For a closed 2d surface 
$M$ this term is defined as an integral of a 3-form over a 3-manifold 
$\Gamma$ with the 2d surface $M$ as its boundary.
If the 2d surface has a boundary the unambiguous definition of this term is problematic.

We need the condition $H^3(G)=0$ on the group $G$ in order to define the
topological term in the WZW model in terms of a globally
well-defined 2-form $w_2$ such that $dw_2(g)=w_3(g)=\tr(dg g^{-1})^3$
(since $w_3$ is a closed 3-form, $dw_3=0$, such $w_2$ always exists
locally). We write this formally as $w_2(g)=d^{-1}w_3(g)$. 
If
$H^3(G) =\ensemble Z$ 
there is no such globally well-defined $w_2$, but
$\Gamma(g)=\int_M w_2$ is still globally well-defined modulo
$\ensemble Z$ as long as $M$ has no boundaries. If $M=\Sigma$ has a
boundary, one needs the condition $H^3(G)=0$ in order to define
$\int_{\Sigma} w_2(g)$ for an arbitrary map $g:\Sigma\to G$. This is
the case, for instance, for $SL(2,\ensemble R)$ which we will now
consider. However, even in this case $\Gamma(g)$ is not unique since
any $w_2'$ that differs from $w_2$ by an exact 2-form,
\eqn\omegatwo{w'_2=w_2(g)+d\beta(g),}
leads to the same $w_3$. In general  $d\beta$
is closed but not necessarily exact. Thus, the action
$\Gamma(g)$ is defined by the 3-form $w_3$ up to an ambiguity that
comes from the 1-form $\beta$, which contributes to the action only
through a boundary term
\eqn\amb{\Gamma^{\beta}(g)=\int_{\Sigma} w_2(g) + \int_{\partial \Sigma} \beta(g^b).}
We denote by $g^b$ the restriction of $g$ to the boundary. If $\beta$ is not well-defined globally, $\int_{\Sigma} d\beta$ still makes sense and 
depends only on $g^b$ since for two different continuations of $g^b$ into the bulk the difference is
$\int_{S^2}d\beta=0 \,\; {\rm mod}\,\; \ensemble Z$.  

For  
$SL(2,\ensemble R)$,  \amb\ can serve as definition of a class
of WZW actions together with the standard kinetic term

\eqn\stand{I_{WZW}=\frac{\kappa}{4\pi i}\int_{\Sigma} \tr\, ({\partial}_{\mu}g g^{-1})^2+\frac{\kappa}{4\pi i}\Gamma^{\beta}(g).}
One expects the theory to be exactly conformally invariant for particular choices
of the boundary term $\int_{\partial \Sigma} \beta$. Classifying such 1-forms $\beta$ is
an interesting question, in particular, in view of
solutions to the quantum conformality condition
$T=\bar T$ on the boundary which do not reduce to the
condition that $g^b$ belongs to a fixed conjugacy class, which in turn follows from
the equations for the currents $J=\bar J$ on the boundary. The latter constraint is, in fact, stronger than the conformality condition. 

Let us now see how the procedure described
for free scalar field in the previous section is modified in this case. From 
$dw_2=w_3$  
it follows immediately on the level of differential forms that
\eqn\difffor{\gamma(g_1,g_2) \equiv w_2(g_1g_2)-w_2(g_1)-w_2(g_2)+\tr\, g_1^{-1}dg_1 dg_2 g_2^{-1}. }
is a well-defined closed 2-form. We note in passing that \difffor\ is closed without restriction to $H^3(G)=0$.
Furthermore, $\gamma$ defines a 2-cocycle on the loop group $\hat{LG}$. Indeed, if
we integrate the closed 2-form \difffor\ over the disk with boundary $S^1$, we get
$\alpha_2(g^b_1,g^b_2)=
\int_{D}\gamma(g_1,g_2)$,
where $g^b$ is the restriction of $g$ to the
boundary and this $\alpha_2$ satisfies the cocycle condition. To see that
$\alpha_2$ only depends on the boundary data of $g_1$ and $g_2$,
we note that for two different extension
$g_i^+$ and $g_i^-$ 
of $g_i^b$
the difference
\eqn\EQNdiff{
    \int_{D^+} \gamma - \int_{D^-} \gamma=
    \int_{S^2} \gamma = 0 \,\; {\rm mod}\,\; \ensemble Z
}
as a consequence of \difffor.
Since $g_i^+$ and $g_i^-$ are the same on the boundary and are otherwise independent,
the result follows. The fact that $\alpha_2$ satisfies cocycle condition can be checked
by direct algebraic computation using \difffor\ (see also \refs{\fs\michI\michII{--}\lmns}).

To continue we will use the following decomposition (motivated by the free field example
in the previous section) for a generic map from the disk $\Sigma$ to the group $G$:
\eqn\decom{g(z,\bz)=g_0(z,\bz)k(z,\bz); \quad \quad {g_0}|_{\partial \Sigma}=1; \quad \quad
k|_{{\partial \Sigma}}=f(\theta),}
so
$g_0$ describes the D-instanton and $k$ is purely defined by the boundary data
$f(\theta): S^1 \rightarrow G$. We will give a concrete definition of $k$ below. 
For $H^3(G)=0$ each 2-form appearing on the rhs of \difffor\
is separately well-defined,
so that
\eqn\coc{\int_{\Sigma}w_2(g_0k)=\int_{\Sigma}w_2(g_0)+\int_{\Sigma}w_2(k)-
\int_{\Sigma}\tr\, g_0^{-1}dg_0 dk k^{-1}} mod $\ensemble Z$ since the
2-cocycle 
$\alpha_2(g^b_1\equiv 1,g^b_2\equiv f)=0 $ 
mod $\ensemble
Z$ \PressleyQK. Combined with the kinetic term in \amb\ this leads to
the expression 
\eqn\actionsl{I_{WZW}(g)=I_{WZW}(g_0)+\frac{\kappa}{4\pi i}\int_{\Sigma}
\tr\, (\partial_{\mu}k k^{-1})^2+\frac{\kappa}{4\pi i}\Gamma^{\beta}(k)+ \frac{\kappa}{2\pi i}\int_{\Sigma}\tr\,
g_0^{-1}{\bar \partial} g_0 \partial k k^{-1}.}
This action is
well-defined though the theory depends on the 1-form $\beta$ through
the boundary integral $\int_{S^1}\beta(f)$ in $\Gamma^{\beta}(k)$.

In order to proceed we will now specify the extension $k(z, \bz)$
of the boundary data $f(\theta)$ by
solving the Riemann-Hilbert problem
for $f(\theta)$. This means we decompose $f(\theta)$ as
\eqn\rh{f(\theta)=h_+(\theta)h_-(\theta),}
where $h_+$ can be holomorphically continued to $h(z)$ into the
disk $\Sigma$ and $h_-$ anti-holomorphically to ${\bar h}(\bz)$.
Thus, we have for $k(z,\bz)$
\eqn\kzz{k(z,\bz)=h(z){\bar h}(\bz) \quad \quad \quad
}
Here $h$ and $\bar h$ are fields on the complexified group\foot{The factors $h$ and $\bar h$ can be constructed by solving the equation of motion in Minkowski signature, that is, $k(\sigma^+,\sigma^-)=h(\sigma^+){\bar h}(\sigma^-)$ where $h$ and $\bar h$ are independent functions and then define $h(z)$ and ${\bar h}(\bz)$ by analytic continuation.}.
This $k(z,\bz)$ solves the WZW equations of motion and, together with $g_0$,
gives an unique decomposition of an arbitrary field
$g=g_0k$ on the disk. We will take \actionsl\
with this decomposition as definition of the WZW model on the
disk for arbitrary boundary fields $f(\theta)$ taking values in the
group manifold.
In background independent open string field theory we are instructed to
integrate over $f(\theta)$.
As we emphasized above this WZW theory on the disk depends on
the 1-form $\beta$ on the boundary, and since this 1-form is completely
arbitrary we include it in the definition
of the boundary perturbation in BSFT.
We do not specify for which $\beta$ this theory is conformal -- this
is a good question and the only comment we will make is that
the string field theory action is one candidate
for the solution -- its critical points correspond to conformal boundary interactions
parametrized by $\beta(f)$.
We conclude that the WZW theory on the disk for the case $H^3(G)=0$
is given by the action \actionsl\ with the definitions \rh, \kzz\ and \decom.


\subsec{Bulk-boundary factorization}

Unlike for the free field case, in the classical WZW action \actionsl\
the boundary field $k$ does not decouple from the bulk fields $g_0$ 
on the level of the classical action.
The interaction between these two fields is given by 
\eqn\EQNW{
    \frac{\kappa}{2\pi i}\int_\Sigma d^2z\bar J_{g_0} \P K,
} 
where $\bar J_{g_0}$ is the anti-holomorphic $g_0$-current, and
the holomorphic function $K(z)$ is defined via $\P K = \P k k^\1$ using the fact that $\P k k^\1$ is a holomorphic 1-form. 

Nevertheless, we will show below that this cross term in \actionsl\
between $g_0$ and $f$ (which parametrizes $k$) does not
contribute to the path integral over $g_0$. Concretely we will prove
that the $n$-point function
\eqn\npt{\biggl\langle\Bigl(\int_{\Sigma}\tr\,
g_0^{-1}{\bar\partial} g_0 \partial k k^{-1}\Bigr)^n\biggr\rangle_{g_0}=0.}
Thus for any choice of $\beta$
\eqn\antwzw{\int_{g|_{\partial \Sigma}=
f} D[g]e^{-I_{WZW}(g)}=
\Z_0 e^{-W(f)},}
where
\eqn\wlive{W(f)=\frac{\kappa}{4\pi i}\int_{\Sigma} \tr\, (\partial_{\mu}k k^{-1})^2+\frac{\kappa}{4\pi i}\Gamma^{\beta}(k)}
and
\eqn\znot{\Z_0=\int_{{g_0|}_{\partial \Sigma}=1}D[g_0]e^{-I_{WZW}(g_0)},}
verifying our conjectured factorization in this class of models. This is the main
technical result of this paper

We will now give a qualitative argument for the vanishing 
of the n-point function \npt. The explicit proof of 
this claim is given in
the appendix. Consider the functional integral over $g_0$ 
at fixed
$k$. This is the WZW theory with boundary conditions in the 
identity conjugacy class.  
The correlators of $\bar J$'s are functions with poles in $\bz$,
but no positive powers of $\bz$ occur. These correlators are then
multiplied by functions $\partial K$ which are polynomials of
positive powers of $z$. These products are proportional to a
positive power of $e^{i\t}$ so that the $\t$ integral vanishes as
long as no singularities occur and the $U(1)$-action by $e^{i\t}$ is
unbroken. In the appendix we show that no such singularities
appear\foot{It should be noted that this argument works only because
we integrate over the disk. One-dimensional
integrals  of such perturbations over the boundary of the disk would give rise to divergences
\BachasSY.}. A similar situation appears e.g. in Kontsevich's work
\KontsevichVB\
in proving the
vanishing theorem for deformation quantization.

The crucial property for these arguments to work is that the
$n$-point functions of the antichiral currents $\bar J$ on the disk
are functions of $\bz$ only. In general one might expect
interactions of the currents with their images. This would generate
terms which behave singular at the boundary. But in this particular
case no such terms appear, and this is due to the following
argument: It is important that in this $n$-point function
only chiral bulk fields are involved in the WZW theory with
$g_0=1$ at the boundary (for the trivial conjugacy class),
i.e. we are interested in the expectation values $\<\bar
J(\bz_1)\cdots\bar J(\bz_p)\>_{D}$ with Dirichlet boundary
conditions. This amplitude has an equivalent representation in terms
of a Dirichlet boundary state $|B_D\>$. The explicit construction of
$|B_D\>$ is not needed. We merely need to assume that such a state
exists. Then the expectation value can be written as an unnormalized
correlation function $\<0|\bar J(\bz_1)\cdots\bar J(\bz_p)|B_D\>$.
Expanding the currents in modes we get 
\eqn\EQNcurrentI{
        \sum_{n_1\cdots n_p}\bz_1^{n_1}\cdots \bz_p^{n_p}\<0|\bar
j_{n_1}\cdots\bar j_{n_p}
        |B_D\>.
}
All $\bar j_n$ with $n\le 0$ annihilate on the vacuum, thus the
expression contains only terms with $n>0$. 
The boundary state is
defined by $\bar J d\bz|B_D\> = Jdz|B_D\>$, thus it maps $\bar j_n$
to $j_{-n}$. Since the holomorphic and anti-holomorphic currents
commute, the $j_{-n}$ can be moved all the way to the left to act on
the vacuum, which it annihilates. This then implies that bulk normal
ordered monomials of chiral operators have a vanishing expectation
value also for Dirichlet boundary conditions. For 
the ordinary product of antiholomorphic currents we then conclude that 
\eqn\EQNcurrentII{
    \<0|\bar J(\bz_1)\cdots\bar J(\bz_p)|B_D\> \propto
        \<0|\bar J(\bz_1)\cdots\bar J(\bz_p)|0\>,
}
that is, the boundary state enters only in the normalization.
Thus the only singularities are those of coinciding  
$\bar J$'s,
which can then be treated in the manner described above.

To summarize, this line of argument shows that, although the bulk and boundary fields to not decouple in the classical action, the partition
function is independent of the interaction term 
$\int \bar J_{g_0}\partial K$ 
to any
order in perturbation theory. 
Thus the boundary
degrees of freedom decouple from the bulk and the partition function
factorizes. To complete the argument we note that the translation
invariance of the functional Haar measure $D[g]$ implies that no
Jacobian occurs when integrating out the bulk fields $g_0$. Thus
\eqn\EQNconclIII{
    \int_{g|_{\partial\Sigma}=f} D[g] e^{-I_{WZW}(g)} = \Z_0 e^{-W(f)},}
where
\eqn\EQNconclIV{
    W(f) = \frac{\kappa}{4\pi i}\int_\Sigma\tr\,(\partial_\mu kk^{-1})^2 + \frac{\kappa}{4\pi i}\Gamma^\beta(k).
    }
An immediate consequence of the above result is that the boundary partition
function on a group manifold is related to the flat space partition
function by a non-local boundary deformation in agreement with the
correspondence stated in the introduction.


\subsec{Groups with $H^3(G)\neq 0$ }

Let us now turn to the case when $H^3(G)=\ensemble Z$ 
such as $G=SU(N)$.
In this case $w_2$ is not globally defined 
(it is ill-defined on a high codimension
submanifold of the target, which is just a point for the case of the $SU(2) \cong S^3$ group manifold).
Let us follow the arguments for the $H^3(G)=0$ case and see where the problems show up. In order to reduce
the action to the form \actionsl\ we use the decomposition $g=g_0k$, with $k$ as in
\rh. 
Since \difffor\ is still well-defined we can formally arrive at the equation \coc.
In particular, the non-trivial 2-cocycle $\alpha_2(g_0^b,f)$ after formula \coc\ is 
again zero for $g^b_0=1$. 
There is no
problem to globally define the first
and the last term on the rhs of \coc. The difficulty resides in the second term $w_2(k)$.
Thus the problem is with the definition of the WZW action on solutions of the 
classical equations of motion 
${\bar \partial}(\partial k k^{-1})=0$, with $k|_{\partial\Sigma}=f(\theta)$, where $f$ is arbitrary. 
In this case the classical Lagrangian turnes out to be
not a function anymore \refs{\GawedzkiYE,\GawedzkiBQ}. 
However, this might be expected, because
a path integral with boundary conditions defines
a wave-function, which corresponds to a 
section of some bundle.
Note that although the action is ambiguous 
the equations of motion derived from it are well-defined.

Recall that the reason we want to consider boundary conditions which are not in a conjugacy class 
is that according to the philosophy of BSFT one has to integrate over all degrees of freedom 
including the boundary fields with boundary interactions parametrized by the 1-form $\beta$. From  the expression  \EBSFT\ for the string field theory action, it follows that 
it is the space-time action \EBSFT\ that needs to be well-defined
for boundary deformations and not the world-sheet classical action 
$W(f)=\frac{\kappa}{4\pi i}\int\tr\,(\partial_\mu kk^{-1})^2 + \frac{\kappa}{4\pi i}\Gamma^\beta(k)$. 
That is, an integral over boundary maps  
\eqn\ZBB{\eqalign{
    \Z^\bdry &= \Z/\Z_0 \cr &=\int 
	D[f]e^{-W(f)}\,,
}}
shall be well-defined, where $D[f]$ is a Haar
measure for $k$ written in terms of $f$ after expressing $k$ via the 
solution of the Riemann-Hilbert problem described above. 
Even if this path integral diverges ultimately, it is the combination entering in \EBSFT\
that shall lead to a well-defined space-time action.

Since there are infinitely
many choices for $\beta$ one would like to classify them according the conformality condition for
the corresponding quantum theory. As we mentioned for $SL(2,\ensemble R)$, this is exactly
the question that background independent open string field theory studies.

For $H^3(G)\neq 0$, one way to remove the topological obstruction in defining
$\Gamma^\beta(k)$, is by deleting a high codimension submanifold in $G$ and
repeating \difffor\ for $g_1=g_0$ and $g_2=k$.
Since these
relations are algebraic we still can safely derive the formal relation \coc. 
One might then suggest that in this case a $d\beta$ can be found, so that
the integral over boundary fields is still well-defined 
as mentioned above (with appropriate regularization procedure).
We recall that a similar situation appears for the
analogous 
quantum-mechanical problem for trajectories with
boundaries in a compact phase-space (associated with coadjoint orbits, and related), 
where the classical action on the world-line
is ill-defined due to non-trivial $H^2$ of the phase space though the path integral can be properly defined
in order to get a correct wave-function\foot{Here, $w_3$ is the Kirillov-Kostant symplectic 2-form.
For a given
$w_2$ (which is a 1-form then) an appropriate $d\beta$ (which also is a 1-form then)
can be found
such that the path integral for an open trajectory, where the boundaries are points now,
gives a matrix element in an irreducible representation of the compact group.
}
\AFS.
In short --
although the action is ill-defined on high codimension submanifolds the path integral
on the manifold with boundary still gives a well-defined and 
correct ``wave-function" (matrix element).
According to \refs{\orbitI, \orbitII}
our current problem is an infinite-dimensional version of the quantum mechanical problem.
We believe that the same is true for the family of 2d field theories related
to WZW models on the disk for group manifolds with non-trivial $H^3$.

Critical points of the string field
theory action \EBSFT\ are supposed to lead to well-defined 
conformal boundary conditions and well-defined $\Z^\bdry=\Z/\Z_0$,
which is the value of the space-time action on-shell according to \EBSFT\ 
(these boundary interactions, in particular, do contain the
restriction to conjugacy classes as a sub-set of the conformal conditions).

So at the moment
we simply assume that \stand\ be given via \actionsl\
for all groups including those with $H^3(G) \neq 0$ (as we mentioned for $SL(2,\ensemble R)$,
everything is properly defined in \actionsl\ and this case is very intersting
on its own right) and define
the string field theory action via standard methods.

At this point a comment about the measure $D[k]$ is in order. If we
start with the Haar measure for $g$, the natural measure for $k$
comes out to be the functional Haar measure for $k(z,\bz) = h(z)\bar
h(\bz)$. Note, however, that $k$ is uniquely determined in terms of
the boundary data. When pulling back $D[k]$ to the
boundary a Jacobian occurs and introduces a further non-locality in
the boundary interaction. So the total non-local boundary
deformation resulting from a shift in the closed string background
is given by $W(f) = \frac{\kappa}{4\pi i}\int_\Sigma\tr\,(\partial_\mu kk^{-1})^2 
+ \frac{\kappa}{4\pi i}\Gamma^\beta(k)$ plus the Jacobian generated.
In the next section
we will give an illustration by considering the large radius limit
of the $SU(2)$ model.

\newsec{Illustrative example and final remarks}
\seclab\SecExample

As we have already mentioned there is no unique way to fix $\beta$
since each choice 
is related to a choice of boundary
interactions. In principle $\b$'s corresponding to conformal open string backgrounds can be constructed order by order in perturbation theory imposing scale invariance at each order. 
As starting point we take the action
\eqn\EQPW{
    S_\Sigma(g_0k) = S_\Sigma(g_0) +
    \frac{\kappa}{2\pi i}\int_{\Sigma}\tr\,g_0^{-1}{\bar \partial} g_0 \partial k k^{-1}
     + \frac{\k}{4\pi i}\int \tr\,k^\1\P kk^\1\bP k. }
In fact the last term can be written in terms of the boundary data $f(\theta)$
(via its decomposition \rh) by using the fact that $h$ and $\bar h$ are
holomorphic and thus writing $\partial h h ^{-1}=\partial K^+(f)$ and ${\bar
h}^{-1}
\bar\partial {\bar h}=\bar\partial K^-(f)$. This leads to an expression
in terms of the Hilbert transform $\oint d{\theta} K H(K)$, where $K=K^++K^-$
(see \hil). This form makes the dependence on
$f$ more explicite.
In the ``Abelian limit'' the latter becomes exactly the formula for the boundary
action we derived for free fields.

We now take \EQPW\ as action
for an $SU(2)$ WZW model and consider the first order perturbation in $1/\sqrt{\kappa}$. That is we use the
parametrization $k = \exp i \frac{X_b^i\s^i}{\sqrt{\k}}$, where $\sigma^i$ are the Pauli matrices. These
coordinates have a simple interpretation in the large-$\k$-limit.
They become the usual flat coordinates and the boundary action (last term in \EQPW) leads upon integration over the boundary fields to a space-filling brane (Neumann boundary conditions in all directions). 
However, we do not expect this boundary interaction to be conformal for finite $\kappa$. 
This is because the quantum mechanical propagator $\frac{1}{\P_n}$ of the boundary
theory is not standard and needs to be renormalized. 
The boundary fields $X_b$
can be re-expressed in terms of holomorphic and antiholomorphic 
continuations of the boundary 
data $f^i(\theta)$ to $f^i(z)$ and $\bar f^i(z)$,
\eqn\EQNsuI{
    X_b^i = f^i + \bar f^i
    - \frac{1}{\sqrt{\k}}
    \epsilon^i_{jk}f^j\bar f^k.
}
Here the large-$\k$-limit  has been taken, including only the first
correction. Writing the boundary interaction in terms of these
fields we get
\eqn\EQNsuII{
    W(f) \sim \oint \Bigl\{f^i\bP \bar f_i +
\frac{1}{\sqrt{\k}}\epsilon^{ijk} f_i\bar f_j\P f_k
        - \frac{1}{\sqrt{\k}}\epsilon^{ijk}f_i\bar f_j \bP\bar f_k
    \Bigr\},
}
which is a non-local function of the flat space  boundary fields
$f^i(z,\bar z)=f^i(z)+\bar f^i(\bar z)$. Next we consider the Jacobian from expressing
the boundary Haar measure D[k] in terms of the flat space measure.
To first order the correction to the measure is then given by the
formal expression
\eqn\EQNsuIII{
    -\frac{1}{\sqrt{\k}}\tr \frac{\d}{(\d f^l, \d\bar f^l)}\epsilon^{ijk}
\left [f^j \bar f^k\right ]^\pm,
}
where $\pm$ stands for decomposition in holomorphic and
anti-holomorphic modes. Variation will always produce factors
$\d^{jl}$ or $\d^{kl}$. As the trace also includes a contraction of
the indices $i$ and $l$, the first correction to the measure
vanishes. Therefore the first correction to the boundary partition
function is \eqn\EQNsutwocorrection{
    \Z^{bdry}\left(t^i\right) = \VEv{1 -
\frac{1}{\sqrt{\k}}\oint\epsilon^{ijk} f_i\bar f_j\P f_k
        + \frac{1}{\sqrt{\k}}\oint\epsilon^{ijk}f_i\bar f_j \bP\bar
f_k}_{I^\bdry(t^i)}. }
The expectation value is taken with respect to
the canonical action in flat space $\frac{1}{2}\oint f\P_n
f$ and a boundary interaction $I^\bdry$, using the flat measure.
This expression can now be used to determine the first order contribution
of $\beta$ by imposing scale invariance.

To summarize, in this paper we have given evidence for the conjecture that within the 
framework of BSFT different closed string $\sigma$-model backgrounds can be 
equivalently described in terms of non-local open string backgrounds. 
This serves as a test for the idea that closed string degrees of freedom are 
indeed contained in the classical open string field theory.

\bigskip

{\bf Acknowledgements:}
We would like to thank
A. Yu. Alekseev,
C. Bachas,
J. Figueroa-O'Farrill,
A. A. Gerasimov,
M. Gaberdiel,
B. Jurco,
M. Kontsevich,
N. Nekrasov,
A. Recknagel,
V. Schomerus,
H. Steinacker
and
L. Takhtajan
for stimulating discussions. 
This work is supported by RTN under contract 005104 ForcesUniverse.
The research of S. Sh. also is in part supported
by Enterprise Ireland Basic Research Grant.

\vfill\supereject

\newsec{Appendix}
\seclab\SecAppendix

In this appendix we give an explicit proof of the claim in section 3.1 that that 
\EQNW\ does not contribute to the path integral over the bulk field $g_0$.
We choose coordinates $z=\r e^{i\t}$ on the disk ($|z|\le 1$). The
operator $\exp -\int \bar J_{g_0}\partial K$ is expanded as $\sum (n!)^\1 (-1)^n I_n$, where
\eqn\EQNappI{\eqalign{
    I_n &\equiv \int d^2 z_1 \P_1 K(z_1) \cdots \int d^2 z_n \P_n K(z_n)
{\cal A}_n(\bar z_1, \dots, \bar z_n) \cr
    {\cal A}_n &\equiv \vev{\bar J_{g_0}(\bz_1)\cdots\bar J_{g_0}(\bz_n)}.
    }
}
Here $J_{g_0}=g_0^{-1}\bar\partial g_0$ is the anti-holomorphic bulk current. 
The basic ingredient for computing the integral \EQNappI\ is the OPE of the anti-holomorphic currents
\eqn\EQNappII{
    \bar J^a(\bz_1)\bar J^b(\bz_2) \sim \frac{\k \d^{ab}}{(\bz_1-\bz_2)^2}
+ \frac{if^{abc}}{\bz_1-\bz_2}
    \bar J^c(\bz_2),
}
where $\bar J = \k^\1 \bar J^a T^a$, $T^a$ are the generators of the
algebra, $f^{abc}$ the structure constants and $\d^{ab}$ the Cartan
metric. But we will see that the calculation does not depend on
details like symmetry structures of the group.

As general strategy we evaluate the indefinite integrals in
order to treat the singularities correctly. The result is then shown
to be a regular function of all variables, so that the boundaries
can be inserted and no singularities occur.

It is clear that the one-point function vanishes, $I_1$=0.
The two-point function is more involved since the Wick theorem does
not hold and there are self-interactions of the currents. The
amplitude is
\eqn\EQNappIII{
    {\cal A}_2(\bar z_1, \bar z_2)
    = \VEv{ \frac{k\d^{ab}}{(\bar z_1 - \bar z_2)^2} + \frac{ i
f^{abc}}{\bar z_1-\bar z_2} \bar J^c(\bar z_2)  } T^a T^b
    \propto \frac{1}{(\bar z_1 - \bar z_2)^2}.
}
We expand the holomorphic field as
\eqn\EQNappIV{
    \P K(z) = \sum_{m>0} m K_m z^{m-1}.
}
Thus, $I_2$ consists of (a sum of) terms
\eqn\EQNappV{
    m_1m_2\oint d\t_1 e^{i(m_1-1)\t_1}\oint d \t_2
e^{i(m_2-1)\t_2}\int_0^1 d\r_1  \int_0^1 d\r_2
\frac{\r_1^{m_1}\r_2^{m_2}e^{2i\t_1}}{(\r_1-\r_2e^{-i(\t_2-\t_1)})^2}.
}
The structure becomes more obvious when a relative boundary
coordinate $\t=\t_2-\t_1$ is introduced,
\eqn\EQNappVI{
    m_1m_2\oint d\t_1 e^{i(m_1+m_2)\t_1}\oint d \t e^{i(m_2-1)\t}\int_0^1
d\r_1  \int_0^1 d\r_2
        \frac{\r_1^{m_1}\r_2^{m_2}}{(\r_1-\r_2e^{-i\t})^2}.
}
As $m_i\ge 1$ the $\t_1$-integral makes the whole term vanish as
long as the remaining integrals are not divergent. The $m_i$ are set
to $1$, because higher powers of $\r_i$ will, at best, smoothen the
singularities. We conduct the $\r_1$-integral and the relevant part
becomes
\eqn\EQNAppCalcI{
    \int d\t \int d\r_2
        \left[
        \r_2\ln(\r_1-\r_2e^{-i\t}) -
\frac{\r_2^2e^{-i\t}}{\r_1-\r_2e^{-i\t}}
        \right].
}
The second part of \EQNAppCalcI\ is
\eqn\EQNAppCalcII{
        \eqalign{
            &\int d\t e^{-i\t}\left[
            \2\r_2^2e^{i\t} +
\r_1\r_2e^{2i\t}+\r_1^2e^{3i\t}\ln(\r_1-\r_2e^{-i\t})
        \right] \cr
    &= \r_1^2\int d\t e^{2i\t}\ln(\r_1-\r_2e^{-i\t}) + {\rm regular\; terms}. 
}} 
Conducting the $\t$-integral yields
\eqn\EQNappVII{
	-\frac{i}{2}\left(\r_1^2e^{2i\t}-\r_2^2\right)\ln\left(\r_1e^{i\t}-\r_2\right)
		+ {\rm regular\; terms}.
}
which is non-singular in all variables. Therefore the whole
expression is non-divergent and vanishes finally under the
$\t_1$-integral.

The first part of \EQNAppCalcI\ is, after $\r_2$-integration,
\eqn\EQNappVIII{
    \2\int d\t \left[
        \r_2^2\ln(\r_1-\r_2e^{-i\t}) + e^{-i\t}\int
d\r_2\frac{\r_2^2}{\r_1-\r_2e^{-i\t}}
        \right].
}
The whole expression becomes, using the result from \EQNAppCalcII,
\eqn\EQNappIX{
    -\2\int d\t
        e^{2i\t}(\r_1^2-\r_2^2e^{-2i\t})\ln(\r_1-\r_2e^{-i\t})
            + {\rm regular\; terms}.
}
This term is regular even without $\t$-integration. Therefore all
terms are finite and finally vanish under the $\t_1$-integral. Thus
$I_2=0$.

The three-point-amplitude is proportional to
\eqn\EQNappX{
    {\cal A}_3(\bar z_1, \bar z_2, \bar z_3)
    \propto \frac{1}{ (\bar z_1- \bar z_2)(\bar z_1-\bar
z_3)(\bar z_2 - \bar z_3)}.
}
$I_3$ contains terms of the form
\eqn\EQNappXI{
     \int d\r_1 d\t_1 \cdots d\r_3 d\t_3
    \frac{m_1m_2m_3\r_1\r_2\r_3}
{(\r_1e^{-i\t_1}-\r_2e^{-i\t_2})(\r_2e^{-i\t_2}-\r_3e^{-i\t_3})(\r_1e^{-i\t_1}-\r_3e^{-i\t_3})}.
}
Again we set $m_i=1$ in order to single out the most singular part.
The indefinite integration over $\r_1$ gives \eqn\EQNAppCalcIII{
    \eqalign{
    &\oint d\t_1\int d\r_1 \frac{\r_1\r_2\r_3}
{(\r_1e^{-i\t_1}-\r_2e^{-i\t_2})(\r_2e^{-i\t_2}-\r_3e^{-i\t_3})(\r_1e^{-i\t_1}-\r_3e^{-i\t_3})}\cr
    &=\frac{\r_2^2e^{-i\t_2}\r_3}{(\r_2e^{-i\t_2}-\r_3e^{-i\t_3})^2} \oint
d\t_1
        e^{2i\t_1} \ln(\r_1e^{-i\t_1}-\r_2e^{-i\t_2})
         - \left[ \r_2e^{-i\t_2} \leftrightarrow \r_3e^{-i\t_3} \right].
}} Now we conduct the $\t_1$-integral\foot{We multiply the integrand
with $e^{-i\t_1}$, which does not change the degree of divergence.
We could also use the integrand without modifications, but the computation is
slightly longer.} 
\eqn\EQNAppCalcIV{
    \eqalign{
    &\int d\t_1 e^{i\t_1} \ln(\r_1e^{-i\t_1}-\r_2e^{-i\t_2}) \cr
    &=i\r_1\left[
\frac{\r_1e^{-i\t_1}-\r_2e^{-i\t_2}}{\r_1e^{-i\t_1}\r_2e^{-i\t_2}}\ln(\r_1e^{-i\t_1}-\r_2e^{-i\t_2})
        -\frac{e^{i\t_2}}{\r_2}\ln(\r_1e^{-i\t_1}) \right].
}} Restoring the pre-factors from \EQNAppCalcIII\ we see that
\EQNAppCalcIV\ is less singular than
\eqn\EQNappXII{
    \frac{i\r_2\r_3e^{i\t_1}}{{\bar z_{23}}^2}
        \left(
        \bar z_{12}\ln\bar z_{12}
        - \bar z_1\ln\bar z_1
        \right) - \left[ \r_2e^{-i\t_2} \leftrightarrow
\r_3e^{-i\t_3}\right].
}
The expression in the bracket is completely regular. As pre-factor
we recognize the contribution from the 2-point function. Thus, we
conclude that $I_3$ must have the same or a less singular behavior
than $I_2$. Thus, the overall $\t_1$-integration, which is also
present for the three-point function, makes the whole expression
vanish, $I_3=0$.

This argument can now be applied recursively to $n$-point functions.
For the sake of a clear presentation we switch to a rather symbolic
notation. The recursion then works like (modulo some permutations)
\eqn\EQNappXIII{
    \eqalign{
    \int d z_n &\frac{(\cdots)}{(\bz_1-\bz_2)(\bz_2-\bz_3)\cdots
        (\bz_{n-1}-\bz_n)(\bz_n-\bz_1)} \cr
    \propto &\frac{(\cdots)}{(\bz_1-\bz_2)(\bz_2-\bz_3)\cdots
        (\bz_{n-1}-\bz_1)} + {\rm less\; singular\; terms}
    }
}
until one ends up with a three-point amplitude. Thus, all these
indefinite integrals are indeed regular.

Now we argue that in fact all the integrals $I_n$ must vanish. We
extract $e^{i\t_1}$ from each factor $(\bz_i-\bz_j)^\1$ and shift
all the other boundary coordinates $\t_i\to \t_i'=\t_i-\t_1$. This
gives a global factor of $\exp i\sum_{i}m_i\t_1$. The $m_i$ are
always positive, thus the $\t_1$-integration makes the whole
expression vanish. We arrive at the central result of this
calculation:
\eqn\EQNappXIV{
    I_n = 0.
}
The immediate consequence is that the operator 
$\exp -\int \bar J_{g_0}\partial K$ 
is marginal
and therefore the partition function does not depend on it.

\listrefs
\bye